\documentclass[10pt]{iopart}

\usepackage{graphicx}
\usepackage{dcolumn}
\usepackage{bm}
\usepackage{epstopdf}
\usepackage{iopams}

\begin{document}

\title[The application of the weak electric field pulse to measure the pseudo-Stark split by the photon echo beating]
{The application of the weak electric field pulse to measure the pseudo-Stark split  by the photon echo beating}

\author{V.N.Lisin, A.M.Shegeda, V.V.Samartsev}

\address{Zavoisky Physical-Technical Institute, Russian Academy of Sciences, Kazan, 420029 Russia}
\ead{vlisin@kfti.knc.ru}
\vspace{10pt}
\begin{indented}
\item[]November 2014
\end{indented}

\begin{abstract}
New scheme for determining the pseudo-Stark splitting of optical lines
has been proposed and experimentally realized. A pulse of a weak
electric field leads to occurrence of relative phase shifts of the
excited dipoles and, as consequence, to beating of a photon echo
wave form if electric pulse  overlaps in time with echo-pulse. The value
of the pseudo-stark splitting is the inverse period of the beats.
The photon echo beating of the R1-line in Ruby have been observed.
The pseudo-Stark splitting has been determined and it's value have been compared
with the known literary data.
\end{abstract}

\pacs{42.50.Md, 42.65.Vh}

\noindent{\it Keywords}: photon echo, phase control, low electric field, pseudo-Stark splitting,
Cr$^{3+}$, Ruby

%

\ioptwocol

\section{Introduction}

Paper purpose is to observe the  the photon echo beating of the R1-line in Ruby,
the pulse applying weak electric field, and measure  the pseudo-Stark splitting.

Photon echo is coherent radiation of medium in form of short pulse,
caused by restoration of phase of separate radiators after the
change of a sign on relative frequency of radiators. If perturbation splits optical line,
for example, on two lines, this means that the frequency of transitions in two groups
of radiators are shifted by different values. If a pulse of the perturbation overlaps
in time with echo-pulse then the echo waveform changes and photon echo beats must be.
The photon echo beats have been observed firstly
in systems in which the Zeeman splitting is realised \cite{1, 2, 3}.
In systems, in which the pseudo-Stark effect is manifested, the line splitting in electric
field  are due to the different Stark shifts at different lattice sites (as you cay see from figure 1). It is well known
that in Ruby there are two kinds of chromium sites, A and B \cite{4}.
Transformation of the A site into the B site is only possible
by symmetry operations involving inversion around the chromium ion, while transformation
of A to A or B to B is achieved solely by translation or by both translation and rotation around
the optic axis. The A and B sites are energetically equivalent in the absence of an electric
field. The shifts $\delta \nu$ of transition frequencies of A and B ions in the electric field equal in magnitude but differ in sign and the optical line splits:

\begin{equation}
\label{eq1}
\begin{array}{l}
\delta \nu=\pm Z/2,  \\
Z=2 {\partial \nu /\partial E} \cdot E,
\end{array}
\end{equation}

\noindent Here  $Z$ is optical line splitting, $E$ is the component of the external electric field along the crystal axis $C_3$ in Ruby, the parameter $\partial \nu /\partial E$ is proportional to the difference of the values of diagonal matrix elements of the operator of the electric dipole moment in the ground and excited states of the R1 line .
Relative phase $\varphi$ of these dipole groups is not equal to zero as result:
\begin{equation}
\label{eq2}
\varphi(t) = 2\pi\int\limits_{t_{0}} ^{t} {Z(t')} dt'.
\end{equation}

\noindent Echo intensity oscillates versus phase \cite{1,2,3} :

\begin{equation}
\label{eq3}
  I\left (t \right) = I_{0}\langle
  cos\left(\varphi ( t)/2 \right)\rangle ^{2},
\end{equation}

\noindent where $I_{0}(t)$  is intensity of an echo when a electric pulse (EP) off,
$t_0$ is time of the beginning of action of a EP, $Z$ is pseudo-Stark splitting of
optical line (twice the value of the pseudo-stark shift), $\langle ... \rangle$ denotes
averaging by optically excited volume.

\begin{figure}

\includegraphics[width=8cm]{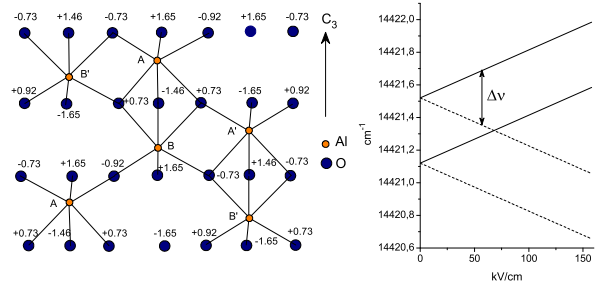}
\caption{\label{fig:stark} At the left  Ruby structure \cite{4} is shown. At the right the splitting of a
R1-line of absorption of a Ruby is conditionally shown in external constant electric field in size
to 160 kV/cm \cite{5}. Continuous and dashed lines belong to ions of chrome of different
type (A and B ions).}
\end{figure}

 Let's pass to consideration of experimental technics.

\section{Experimental conditions}
We observed the photon echo in the backward mode on the R1-line of the
(Al$_{2}$O$_{3}$:Cr$^{3+})$  at temperature of 2 K.
The laser pulses with duration of 13 ns have
been generated by an oxazine--1 dye laser. The spectral FWHM is
 about 6 GHz. Delay time between the laser pulses is less than or equal to 64 ns.
Polarization of the laser pulses is  linear. The photon echo
signals were detected on a photomultiplier and fed to a Tektronix
TDS 2022 digital oscilloscope.  All experimental measurements were
controlled with LabVIEW software. The geometry of the experiment
is shown in figure 2. The electric field is obtained by applying a
voltage difference to a pair of electrodes press down to an opposite faces of
the crystal with the resulting electric field parallel to
$C_3$-axis.  Each electrode is thin copper plate with hole  in
center. Hole diameter is 1 mm. Axis X passes through the centers of the holes of the electrodes, directed in parallel to the crystal $C_3$-axis. The origin is in the center between the plates: x=0. The direction of propagation of the laser beams are approximately parallel to the $C_3$-axis. The angle between the
rays of the first and second laser pulses is about 4,5 degrees.
The diameter of the focused beams in sample is 0.15 mm. Dependence of the intersection area of beams versus $x$ is shown in figure 2.

Voltage pulses formed by a generator with avalanche transistors
are applied to the electrodes synchronously with the laser
pulses. The half max voltage pulse duration $\tau$$_{U}$ equal 34
ns and upper flat part duration is 23 ns as you can see in figure
2. The maximum amplitude of a voltage difference is equal 268 V. The voltage
amplitude was varied with a step of 1dB.

The electric field between the electrodes with holes was calculated
by numerical solution of the Poisson equation. We Take into account
that the electrodes are equipotential surfaces
and electric field on the surface of the electrodes directed normal
to the electrode surfaces.

Calculations showed (figure 2) that the value of the component of the
electric field parallel to the $C_3$ axis in the center of the optically excited
volume is equal 0,993 to an infinite flat field of the capacitor
without holes $E_0$:

\begin{equation}
\label{eq4}  E_0=U/d,
\end{equation}
\noindent where $U$ is voltage difference between the electrodes and $d$ is the sample  thickness.

From the calculations it follows that with good accuracy the relation is valid
\begin{equation}
\label{eq5}
  \langle   cos\left(\varphi ( t)/2 \right)\rangle = cos \left(\langle   \varphi ( t)/2 \rangle\right)
\end{equation}

\noindent in determining the times at which the intensity (3) minima are achieved .
 We take into consideration dependence of the area of intersection of beams on x
 during the averaging over the optically excited volume. See line 2 in figure2 at the right.
 The average value of the electric field is
 \begin{equation}
\label{eq6}
 \langle   E\rangle = 0.987 U/d =0.987 E_{0}
\end{equation}

\begin{figure}
\includegraphics[width=8cm]{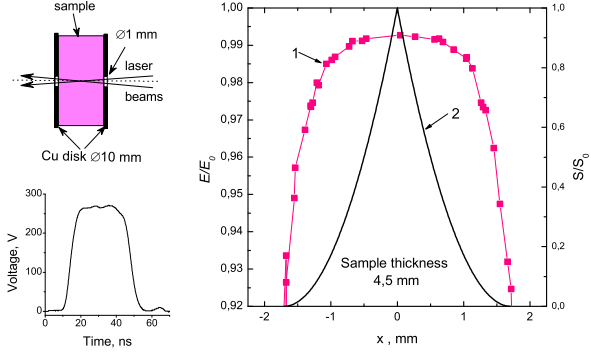}
\caption{\label{fig:exp} At the left above the experiment geometry
is resulted. Arrows represent the rays of the first and second laser pulses, the dot line depicts the $C_{3}$ and $X$-axes.
The condenser is formed by two copper disks in diameter of 10 mm with holes in the centre in diameter of
1 mm. Disks are densely pressed to  the sample. The first and
second laser pulses in diameter ~ 0.15 mm  transit  in holes of
disks at an angle ~ 4,5 degrees rather each other. At the left below
you can see a waveform of a pulse of the voltage difference submitted on
condenser. The calculated values of the electric field $E/E_{0}$ (line 1) and the area of beams intersection $S/S_{0}$
(line 2) versus $x$ are shown at the right. The origin is in the center between the plates: x=0.
$E_{0}$ is defined in (4), and $S_{0}$ is the cross-sectional area of the laser beam $S_{0}=\pi (0.15/2 )^{2}$ mm$^2$ .
}
\end{figure}
Given the fact that the phase change for the period is equal to $2\pi$, make the approximation
\begin{equation}
\label{eq7}
  \langle   \varphi ( t+T) \rangle-\langle   \varphi ( t) \rangle=
  2\pi=2\pi\langle Z \rangle T,
\end{equation}
which is true only on a flat part of the electric pulse form. From (7) you can express pseudo-stark frequency through the reverse period modulation

\begin{equation}
\label{eq8}
 \langle Z \rangle=1/T=2 {\partial \nu /\partial E} \cdot \langle E \rangle
\end{equation}

\noindent and the difference dipole moments using experimentally measured value

\begin{equation}
\label{eq9}
2 \frac{\partial \nu }{\partial E}=
\frac{\partial \langle Z \rangle}{\partial \langle E \rangle}=
\frac{\partial (1/T) }{\partial \langle E \rangle}=
\frac{d}{0.987}\frac{\partial  (1/T)}{\partial U}.
\end{equation}

\section{Results of experiment}
In figure 3 waveforms of observable R1-line echo signals in Ruby samples. You can see also EP waveforms $U$.

\begin{figure}
\includegraphics[width=8cm]{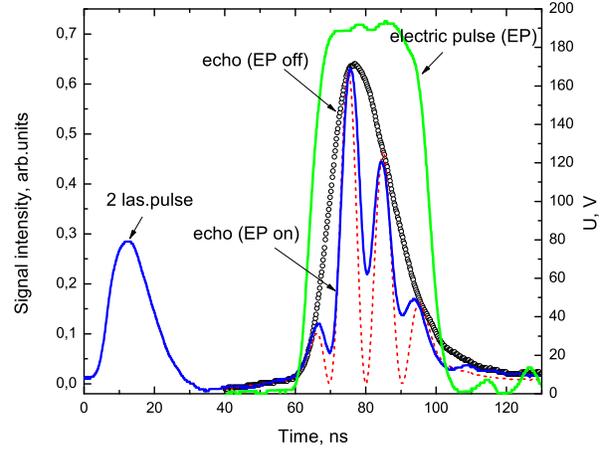}
\caption{\label{fig:echo} Waveforms of electric pulse (EP) and
photon echo  when the EP on and off in Ruby with $^{52}Cr$. Dashed line -
simulation of intensity of a photon echo according to expressions
(\ref{eq3}, \ref{eq2}, \ref{eq7} - \ref{eq9})
}
\end{figure}

From the waveforms of echo signals observed at various times of
the beginning of action of the EP, similar to those shown in
figure 3, time intervals between the nearest minima in the echo
signals were measured and the average value $T$ of the period has
been determined. Also the electric field amplitudes $E_{0}$ and $\langle E \rangle$ in the region of
the top flat time area of voltage pulse has been calculated using equations (\ref{eq4}) and (\ref{eq6}).
The dependence of the inverse average
period 1/$T$ versus the $\langle E \rangle$ has been built. Figure 4 shows these
dependencies for Ruby samples of different thickness and are
enriched for different isotopes of chromium ion $^{52}Cr$ and $^{53}Cr$. We
take into account that echo modulation is absent, i.e. 1/$T$ =0
when $\langle E \rangle$ = 0.

\begin{figure}
\includegraphics[width=8cm]{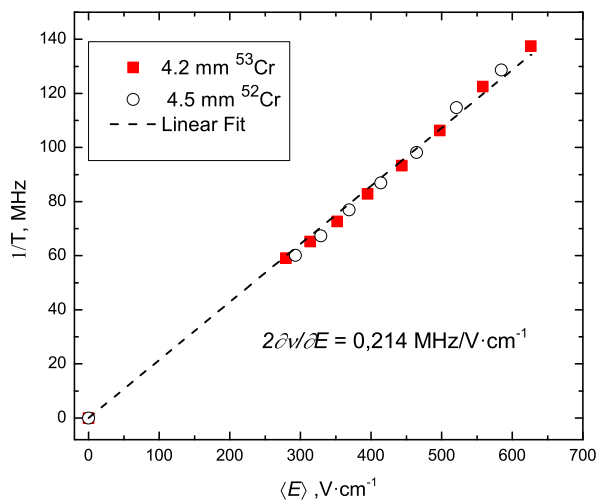}
\caption{\label{fig:time}Inverse average modulation period of the
echo waveform versus $\langle E\rangle$. The $\langle E\rangle$ is
average value of the electric field over the optically excited volume (\ref{eq6}),
where $U$ is amplitude of a voltage pulse for the time in the region of a top flat of the voltage
pulse. }
\end{figure}
From figure 4 the values of following parameters are defined:

\begin{equation}
\label{eq10}
\begin{array}{l}
^{52}Cr, d=4.5 mm  \\
 \partial (1/T) /\partial \langle E\rangle= 0.214 ({MHzV^{-1}cm}), \\\\
^{53}Cr, d=4.2 mm  \\
 \partial (1/T) /\partial \langle E \rangle= 0.214 ( {MHzV^{-1}cm}).   \\
\end{array}
\end{equation}

Taking into account (\ref{eq10}) we have
\begin{equation}
\label{eq11}
2 \partial \nu /\partial E= 0.214({MHzV^{-1}cm}) \}  ^{52}Cr, ^{53}Cr.
\end{equation}

\section{Discussion of results}
The obtained values of the pseudo-stark splitting  parameter are in good agreement
with known literature data.
For comparison, we give the values obtained by other methods in the same units as in (\ref{eq11}):
\begin{equation}
\label{eq12}
2 \partial \nu /\partial E= 0.176 \cite{5}, 0.228\cite{6}, 0.220\pm 0.016 \cite{7}.
\end{equation}

The main reason for the error in the determination
of the pseudo-stark splitting  parameter is unstable laser pulse shape
and, consequently, unstable waveform of the photon echo. In the
ideal case, when the initial shape of the echo signal and the
shape of the EP close to the rectangular, it is enough a single
waveform echo when the EP on to determine the modulation period \emph{T}. Enough to know the
amplitude of the EP and the time between the nearest minima in the
resulting echo response. If the shape of the echo signal is
completely arbitrary, but repetitive from pulse to pulse, knowing
the EP dependence versus time and time $t_{0}$ when the voltage pulse on ,
it is easy to calculate for a given $\partial \nu /\partial E$  the
waveform of the resulting echo response . Comparing the period of
the modulation of the echo signal with the experiment, we can find
the $\partial \nu /\partial E$. In this case, you have two echo
waveforms: with EP off and EP on. If the shape of the echo signal
is completely arbitrary and does not repetitive from pulse to
pulse, it is impossible to know exactly the shape of the echo when
EP off. In this case, to reduce the error we must submit the EP
that creates in the echo waveform as much as possible minima and
then to calculate the average the time between the nearest minima.
Preliminary results were presented at conferences IWQO-2015 \cite{8}.

\section{Conclusions}

By controlling the relative phase of the excited dipoles with weak
perturbations can significantly change the resultant dipole moment
of the system. This can occur in cases where the coherence is
important. In our study we measured the pseudo-stark splitting of optical line
and determined electric parameter $\partial \nu /\partial E$ of the paramagnetic ion.
If the electric parameters are known, it is possible to
determine the amplitude of the EP, using the relation (\ref{eq8}).
For example, if the electric field is created by the pulse changes
the dipole-dipole interaction of the ion with the environment. This
can be used to determine the distance to the centers of the
environment by measuring the waveform of the echo modulation
periods.\\\\

\section{Acknowledgments}
This work was supported by the RAS program 'Fundamental
optical spectroscopy and its applications' and by the RFBR
grants no. 14-02-00041a.\\\\


\end{document}